\documentstyle[prd,aps,epsfig,floats,axodraw]{revtex}
\begin{document}
\draft

\renewcommand{\topfraction}{0.8} 
\newcommand{\beq}{\begin{equation}}
\newcommand{\eeq}{\end{equation}}
\newcommand{\bea}{\begin{eqnarray}}
\newcommand{\eea}{\end{eqnarray}}
\newcommand{\pbar}{\not{\!\partial}}
\newcommand{\dbar}{\not{\!{\!D}}}
\def\lsim{\:\raisebox{-0.75ex}{$\stackrel{\textstyle<}{\sim}$}\:}
\def\gsim{\:\raisebox{-0.75ex}{$\stackrel{\textstyle>}{\sim}$}\:}
\twocolumn[\hsize\textwidth\columnwidth\hsize\csname 
@twocolumnfalse\endcsname

\title{Production of massive stable particles in inflaton decay}   
\author{Rouzbeh Allahverdi and Manuel Drees}
\address{Physik Department, Technische Universit\"at M\"unchen, James Frank
Strasse, D-85748, Garching, 
Germany.}
\date{\today} 
\maketitle

\begin{abstract}

We point out that inflaton decays can be a copious source of stable or
long--lived particles $\chi$ with mass exceeding the reheat
temperature $T_R$ but less than half the inflaton
mass. Once higher order processes are included, this
statement is true for any $\chi$ particle with renormalizable (gauge
or Yukawa) interactions. This contribution to the $\chi$ density often
exceeds the contribution from thermal $\chi$ production, leading
to significantly stronger constraints on model parameters than those
resulting from thermal $\chi$ production alone, particularly in models
containing stable charged particles.

\end{abstract}

\pacs{PACS numbers: 98.80.Cq, 11.30.Pb \hspace*{1.3cm}
TUM--HEP--456/02} 

\vskip2pc]


According to inflationary models \cite{infl}, which were first
considered to address the flatness, isotropy, and monopole problems of
the hot Big Bang model, the Universe has evolved through several
stages.  During inflation, the energy density of the Universe is
dominated by the potential energy of the inflaton and the Universe
experiences a period of superluminal expansion. Immediately after
inflation, coherent oscillations of the inflaton dominate the energy
density of the Universe.  These oscillations eventually decay, and
their energy density is transferred to relativistic particles; this
reheating stage results in a radiation--dominated
Friedmann--Robertson--Walker (FRW) Universe, as in the hot Big Bang
model.

Initially reheating was treated as the perturbative, one particle
decay of the inflaton with decay rate $\Gamma_{\rm d}$, resulting in
$T_{\rm R} \sim {({\Gamma}_{\rm d}{M}_{\rm P})}^{1/2}$ for the reheat
temperature \cite{infl,kt}, where $M_{\rm P} = 2.4 \times 10^{18}$ GeV
is the reduced Planck mass. $T_{\rm R}$ should be low enough so that
the original monopole problem is avoided. Moreover, in many
supersymmetric models ${T}_{\rm R} \leq {10}^{7}-{10}^{9}$ GeV, in
order to avoid gravitino overproduction which would destroy the
success of nucleosynthesis \cite{bbn}.  Later it has been noticed that
the initial stages of inflaton decay might involve non--perturbative
resonance processes \cite{preheat}. They typically lead to a highly
non--thermal distribution of particles, including inflatons with large
momentum \cite{fk}. However, after sufficient red--shifting the energy density
of the Universe would again be dominated by non--relativistic, massive
particles. It is therefore generally believed that an epoch of
(perturbative) reheating from the decay of massive particles (or
coherent field oscillations, which amounts to the same thing) is an
essential ingredient of any potentially realistic cosmological model
\cite{jed}. In what follows we generically call the decaying particle
the ``inflaton'', since we are (almost) sure that inflatons indeed
exist. Note also that in a large class of well--motivated models,
where the inflaton resides in a ``hidden sector'' of a supergravity
theory \cite{rs}, its couplings are suppressed by inverse powers of
$M_{\rm P}$, and hence are so weak that inflaton decays are purely
perturbative. However, it should be clear that our results hold
equally well for any other (late) decaying particle. 

Even before all inflatons decay, their decay products form a plasma
which, upon a very quick thermalization, has the instantaneous
temperature \cite{kt} $ T \sim \left( g_*^{-1/2} H \Gamma_{\rm d}
M^{2}_{\rm P} \right)^{1/4}$, where $H$ is the Hubble parameter and
$g_*$ denotes the number of relativistic degrees of freedom in the
plasma. This temperature reaches its maximum $T_{\rm max}$ soon after
the inflaton field $\phi$ starts to oscillate, which happens for a
Hubble parameter $H_I \leq m_\phi$, with $m_\phi$ being the frequency
of inflaton oscillations about the global minimum of the potential. We
will assume that all inflaton decays can be described by perturbation
theory in a trivial vacuum, which implies $T_{\rm max} <
m_\phi/2$.(The resulting upper bound on $\Gamma_{\rm d}$ also
implies that a vacuum expectation value of the inflaton field does not
induce large masses to the particles to which it couples.) However,
$T_{\rm max}$ can be much larger than $T_{\rm R}$. As long as $T >
T_{\rm R}$ the energy density of the Universe is still dominated by
the (non--relativistic) inflatons that haven't decayed yet. The
Universe remains in this phase as long as $H > \Gamma_{\rm d}$. During
that epoch particles $\chi$ with mass $T_{\rm max} > m_\chi > T_{\rm
R}$ can be produced copiously from the thermal plasma
\cite{ckr,gkr,rosenfeld,ky}. Here we point out that $\chi$ particles
can also be produced directly in inflaton decays. We will show that
the $\chi$ abundance from inflaton decay often exceeds that from
thermal production, even if the branching ratio for $\phi \rightarrow
\chi$ decays is very small.

We begin our argument by pointing out that $T_{\rm max}$ is frequently
well below $m_\phi$. This is important, since thermal production is
obviously only efficient if $m_\chi \lsim T_{\rm max}$, while inflaton
decay can produce pairs of $\chi$ particles as long as $m_\chi <
m_\phi / 2$. For
perturbative inflaton decay thermalization increases the number
density and reduces the mean energy of the decay products.
Complete thermalization (i.e. both chemical and kinetic) therefore
requires $2 \rightarrow N$ reactions, which change the number of
particles, to be in equilibrium. Since the rate for higher order
processes is suppressed by powers of the relevant coupling constant
$\alpha$, the most important reactions are those with $N=3$. These
reactions have recently been studied in Ref.\cite{ds} where the
scattering of two matter fermions with energy $\simeq m_\phi/2$ (from
inflaton decay) to two fermions, plus one gauge boson with typical
energy $E \ll m_\phi$, is considered. The rate for these
reactions can be large due to the $t-$channel pole of the scattering
matrix element, regulated by a cut--off on the exchanged
momentum, naturally taken to be the inverse of the average separation
between two particles in the plasma \cite{ds}. It turns out that the
largest possible $T_{\rm max}$ is given by \cite{ad2}
\beq \label{tmax}
T_{\rm max} \sim T_{\rm R} \left ( \alpha^3 \left( \frac {g_*}{3}
\right)^{1/3} { M_{\rm P} \over m_\phi^{1/3} T_{\rm R}^{2/3} }
\right )^{3/8}.
\eeq 
Even if $m_{\phi}$ is near its upper bound of $\sim
10^{13}$ GeV \cite{cdo}, for a chaotic inflation model, and $T_{\rm
R}$ is around $10^9$ GeV (saturating the gravitino bound) $T_{\rm
max}$ will exceed $T_{\rm R}$ if the coupling $\alpha^3 \gsim
10^{-8}$. This is easily accommodated for particles with gauge
interactions. On the other hand, recall that $T_{\rm max} <
m_{\phi}/2$.  Together with
eq.(\ref{tmax}), taking $\alpha \lsim 0.1$, this gives $T_{\rm max}
\leq 10^{11} \ (10^5)$ GeV for $T_{\rm R} = 10^9 \ (1)$ GeV.  This
implies in particular that there will be no ``wimpzilla'' production
\cite{ckr} from {\em thermalized} inflaton decay products, since in
this case $m_\chi > T_{\rm max}$.

On the other hand, for $m_\chi \lsim 20 T_{\rm R}$ the standard
calculation \cite{kt} of the density of stable relics
applies. Scenarios with $T_{\rm max} \gsim m_{\chi} \gsim 20 T_{\rm
R}$ have only been investigated relatively recently in
refs.\cite{ckr,gkr,rosenfeld,ky}, which studied $\chi$ production from
the thermal plasma with $T > T_{\rm R}$. If the $\chi$ density was
always well below the equilibrium density, one finds
\beq  \label{ssrate1}
\Omega^{\rm therm}_\chi h^2 \sim \left ({200 \over g_*} \right )^{3/2}
\alpha_\chi^2 {\left ({2000 T_{\rm R} \over m_\chi} \right )}^7 .
\eeq
Here $\Omega_\chi$ is the $\chi$ mass density in units of the critical
density and $h$ is the Hubble constant in units of 100
km$/$(s$\cdot$Mpc). We have taken the cross section for $\chi$ pair
production or annihilation to be $\sigma \simeq \alpha_\chi^2 /
m^{2}_{\chi}$. Note that $\Omega_\chi$ is only suppressed by $(T_{\rm
R}/m_\chi)^7$ rather than by $\exp(-m_\chi/T_{\rm R})$. A stable
particle with mass $m_\chi \sim 2000~T_{\rm R} \cdot
\alpha_\chi^{2/7}$ might thus act as the Dark Matter in the Universe
(i.e. $\Omega_\chi \simeq 0.3$). However, eq.(\ref{tmax}) with $\alpha
= 0.05$ implies that $T_{\rm max} \gsim 1000 T_{\rm R}$ is only
possible if $T_{\rm R} < 2 \cdot 10^{-12} M_{\rm
P}$. Eq.(\ref{ssrate1}) is no longer applicable \cite{gkr} if the
coupling $\alpha_\chi$ is so large that $\chi$ reached chemical
equilibrium; however, it can then still be used as an upper bound on
$\Omega_\chi^{\rm therm}$.

We now discuss the direct production of $\chi$ particles in inflaton
decay. (Other mechanisms for nonthermal production of
superheavy particles have been discussed in \cite{hp}.) Most inflatons
decay at $T \simeq T_{\rm R}$; moreover, the
density of $\chi$ particles produced in earlier inflaton decays will
be greatly diluted. Since inflaton decay conserves energy, the density
of inflatons can be estimated as $n_\phi \simeq 0.3 g_* T_R^4 /
m_\phi$. Let us denote the average number of $\chi$ particles
which are produced in each $\phi$ decay by $B(\phi \rightarrow \chi)$. We
translate the $\chi$ density at $T = T_{\rm R}$ into the present
$\chi$ relic density using the relation \cite{kt}
\beq \label{relden}
\Omega_\chi h^2 
= 6.5 \cdot 10^{-7}\, \cdot \, \frac {200} {g_*} \cdot \frac
{m_\chi n_\chi(T_{\rm R}) } {T_{\rm R}^3 T_{\rm now}}.
\eeq
The $\chi$ density from $\phi$ decay is therefore \cite{rosenfeld}:
\beq \label{phidec}
\Omega^{\rm decay}_\chi h^2 \simeq 2 \cdot 10^8 B(\phi \rightarrow
\chi) \frac {m_\chi} {m_\phi} \frac {T_{\rm R}} {1 \ {\rm GeV}}.
\eeq
Eq.(\ref{phidec}) holds if the $\chi$ annihilation rate is smaller
than the Hubble expansion rate at $T \simeq T_{\rm R}$, which requires
\beq \label{nonequil}
\frac{m_\phi} {M_{\rm P}} > 5 B(\phi \rightarrow \chi)
\alpha_\chi^2 \left( \frac {T_R} {m_\chi} \right)^2 \left( \frac {g_*}
{200} \right)^{1/2}.
\eeq
This condition will be satisfied in chaotic inflation models with
$m_\phi \sim 10^{-5} M_{\rm P}$, if $m_\chi$ is large enough to avoid
overclosure from thermal $\chi$ production alone. It might be violated
in models with light inflaton. In that case the true $\chi$ density at
$T_{\rm R}$ can be estimated by equating the annihilation rate with
the expansion rate:
\beq \label{ommax}
\Omega_\chi^{\rm max} \simeq \frac {5 \cdot 10^7} {\alpha_\chi^2}
\frac {m_\chi^3} {(1 \ {\rm GeV}) \cdot M_{\rm P} T_{\rm R}}
\left( \frac {200} {g_*} \right)^{1/2}.
\eeq
This maximal density violates the overclosure constraint $\Omega_\chi
< 1$ badly for the kind of weakly interacting ($\alpha_\chi \lsim
0.1$), massive ($m_\chi \gg T_{\rm R}$ and $m_{\chi} \gsim 1$ TeV)
particles we are interested in. [Eq.(\ref{ommax}) describes
the maximal $\chi$ density if $\chi$ decouples at $T \sim T_R$. It is
not applicable to WIMPs decoupling at $T < T_{\rm R}$.]  For the
remainder of this article we will therefore estimate the $\chi$
density from inflaton decay using eq.(\ref{phidec}).

Our remaining task is to estimate $B(\phi \rightarrow \chi)$. This
quantity is obviously model dependent, so we have to investigate
several scenarios. The first, important special case is where $\chi$
is the lightest supersymmetric particle (LSP). If $m_\phi$ is large
compared to typical visible--sector superparticle masses, $\phi$ will
decay into particles and superparticles with approximately equal
probability. (This statement is true so long as the
superpotential is quadratic or higher in the inflaton superfield
\cite{ad2}.) Moreover, all superparticles will decay into
one $\chi$ particle and some standard particle(s) at a time scale
which is shorter than the superparticle annihilation time scale \cite{kamy}, as
long as $m_\chi > T_{\rm R}$, even if $\alpha_\chi \simeq 0.1$.  As a
result, if $\chi$ is the LSP, then $B(\phi \rightarrow \chi) \simeq
1$, independently of the nature of the LSP.

\setcounter{footnote}{0}
Another possibility is that the inflaton couples to all particles with
more or less equal strength, e.g. through non--renormalizable
interactions. In that case one expects $B(\phi \rightarrow \chi) \sim
1/g_* \sim 1/200$. However, even if $\phi$ has no direct couplings to
$\chi$, the rate (\ref{phidec}) can be large. The key observation is
that $\chi$ can be produced in $\phi$ decays that occur in higher
order in perturbation theory whenever $\chi$ can be produced from
annihilation of particles in the thermal plasma. In most realistic
cases, $\phi \rightarrow f \bar f \chi \bar \chi$
decays will be possible if $\chi$ has electroweak gauge interactions, where
$f$ stands for some gauge non--singlet with tree--level coupling to
$\phi$. A diagram contributing to this decay is shown in Fig.~1. Note
that the part of the diagram describing $\chi \bar \chi$ production is
identical to the diagram describing $\chi \bar \chi \leftrightarrow f
\bar f$ transitions. This leads to the following estimate:
\beq \label{fourbody}
B(\phi \rightarrow \chi)_4 \sim \frac {C_4 \alpha_\chi^2} {96 \pi^3}
\left( 1 - \frac {4 m_\chi^2} {m_\phi^2} \right)^2 \left( 1 - \frac {2
m_\chi} {m_\phi} \right)^{5 \over 2},
\eeq
where $C_4$ is a multiplicity (color) factor. The phase space factors
have been written in a fashion that reproduces the correct behavior
for $m_\chi \rightarrow m_\phi/2$ as well as for $m_\chi \rightarrow
0$. Occasionally one has to go to even higher order in perturbation
theory to produce $\chi$ particles from $\phi$ decays. For example, if
$\chi$ has only strong interactions but $\phi$ only couples to $SU(3)$
singlets, $\chi \bar \chi$ pairs can only be produced in six body
final states, $\phi \rightarrow f \bar f q \bar q \chi \bar \chi$. A
representative diagram can be obtained from the one shown in Fig.~1 by
replacing the $\chi$ lines by quark lines, attaching an additional
virtual gluon to one of the quarks which finally splits into $\chi
\bar \chi$. The branching ratio for such six body decays can be
estimated as
\beq \label{sixbody}
B(\phi \rightarrow \chi)_6 \sim \frac {C_6 \alpha_\chi^2 \alpha_W^2}
{1.1 \cdot 10^7}
\left( 1 - \frac {4 m_\chi^2} {m_\phi^2} \right)^4 \left( 1 - \frac {2
m_\chi} {m_\phi} \right)^{9 \over 2}.
\eeq
Another example where $\chi \bar \chi$ pairs can only be produced in
$\phi$ decays into six body final states occurs if the inflaton only
couples to fields that are singlets under the standard model gauge group,
e.g. right--handed (s)neutrinos $\nu_R$ \cite{kumy}. [Since
$\nu_R$ decays very quickly, $B(\phi \rightarrow \nu_R) \sim 1$ does
not cause any problem.] Since $\nu_R$ only has Yukawa interactions,
the factor $\alpha_W^2$ in eq.(\ref{sixbody}) would have to be
replaced by the combination of Yukawa couplings $\lambda_{\nu_R}^2
\lambda_t^2 / (16 \pi^2)$. If $2m_\chi < m_{\nu_R}$, $\chi \bar \chi$
pairs can already be produced in four body final states from $\nu_R$
decay. The effective $\phi \rightarrow \chi$ branching ratio would
then again be given by eq.(\ref{fourbody}), with $m_\phi$ replaced by
$m_{\nu_R}$ in the kinematical factors.

Finally, in supergravity models there in general exists a coupling
between $\phi$ and either $\chi$ itself or, for fermionic $\chi$, to
its scalar superpartner, of the form $a \left( m_\phi m_\chi/M_{\rm
P} \right) \phi \chi \chi + {\rm h.c.}$ in the scalar potential
\cite{aem}. A 

%
\begin{center}
\SetScale{0.6} \SetOffset(40,40)
\begin{picture}(225,125)(0,0)
\DashLine(0,50)(75,50){5} \Text(0,25)[l]{$\phi$}
\Vertex(75,50){3}
\ArrowLine(125,25)(75,50) \Text(80,16)[h]{$\bar{f}$}
\ArrowLine(75,50)(125,75) \Text(60,50)[t]{$f$}
\Vertex(125,75){3}
\Photon(125,75)(175,100){5}{4} 
\Vertex(175,100){3}
\ArrowLine(125,75)(175,50) \Text(110,30)[h]{$f$}
\ArrowLine(175,100)(225,75) \Text(145,45)[r]{$\chi$}
\ArrowLine(225,125)(175,100) \Text(145,75)[r]{$\bar \chi$}              
\end{picture}
\vspace*{-13mm}

\noindent
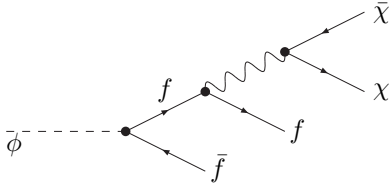
{\bf Fig. 1:}~Sample diagram for $\chi$ production in four-body
inflaton decay. 
\end{center}
\vspace*{3mm}
reasonable estimate for the coupling strength is \cite{aem} $a \sim
\langle \phi \rangle /M_{\rm P}$, unless an $R-$symmetry suppresses
$a$. Assuming that most inflatons decay into other channels, so that
$\Gamma_{\rm d} \sim \sqrt{g_*} T_{\rm R}^2/M_{\rm P}$ remains valid,
this gives
\beq \label{gravbr}
B(\phi \rightarrow \chi) \sim \frac {a^2 m^2_\chi m_\phi} {16 \pi
\sqrt{g_*} M_{\rm P} T^2_{\rm R} } \left( 1 - \frac {4 m_\chi^2}
{m_\phi^2} \right)^{1 \over 2}.
\eeq

The production of $\chi$ particles from inflaton decay will be
important for large $m_\chi$ and large ratio $m_\chi / T_{\rm R}$, but
tends to become less relevant for large ratio $m_\phi / m_\chi$.  Even
if $m_\chi < T_{\rm max}$, $\chi$ production from the thermal plasma
(\ref{ssrate1}) will be subdominant if
\beq \label{compare1}
\frac {B(\phi \rightarrow \chi)} {\alpha_\chi^2} > \left( \frac {100
T_{\rm R}} {m_\chi} \right)^6 \frac {m_\phi} {m_\chi} \frac {1 \ {\rm
TeV}} {m_\chi}.
\eeq
The first factor on the r.h.s. of (\ref{compare1}) must be $\lsim
10^{-6}$ in order to avoid over--production of $\chi$ from thermal
sources alone. Even if $\phi \rightarrow \chi$ decays only occur in
higher orders of perturbation theory, the l.h.s. of (\ref{compare1})
will be of order $10^{-4}$ ($10^{-10}$) for four (six) body final
states, see eqs.(\ref{fourbody}), (\ref{sixbody}); if $\phi
\rightarrow \chi \bar \chi$ decays at tree--level, the l.h.s. of
(\ref{compare1}) will usually be bigger than unity. We thus see that
even for $m_\phi \sim 10^{13}$ GeV, as in chaotic inflation models,
and for $m_\chi \simeq 10^3 T_{\rm R}$, $\chi$ production from decay
will dominate if $m_\chi \gsim 10^7 \ (10^{10})$ GeV for four (six)
body final states. As a second example, consider LSP production in
models with very low reheat temperature. The LSP mass should lie
within a factor of five or so of 200 GeV. Recall that in this case
$B(\phi \rightarrow \chi) = 1$. Taking $\alpha_\chi \sim 0.01$, we see
that $\chi$ production from decay will dominate over production from
the thermal plasma if $m_\phi < 6 \cdot 10^7$ GeV for $T_{\rm R} = 1$
GeV; this statement will be true for all $m_\phi \lsim 10^{13}$ GeV if
$T_{\rm R} \lsim 100$ MeV.

Let us now assume that eq.(\ref{phidec}) indeed gives the dominant
contribution to $\chi$ production in the early Universe, and
investigate the resulting constraints on model parameters.  As well
known, any stable particle must satisfy $\Omega_\chi h^2 < 1$, since
otherwise it would ``overclose'' the Universe. For example, in case of
a neutral LSP with $m_\chi \simeq 200$ GeV, eq.(\ref{phidec}) with
$B(\phi \rightarrow \chi) = 1$ implies $m_\phi / T_{\rm R} > 4 \cdot
10^{10}$. Such a large ratio $m_\phi / T_{\rm R}$ in turn requires
$\Gamma_{\rm d} < 10^{-21} m^2_\phi / M_{\rm P}$, which indicates
that $\phi$ would have to decay through higher dimensional
operators. Of course, this constraint is no longer valid if $\chi$
reaches equilibrium with the plasma at temperatures $\lsim T_{\rm
R}$. 

Another Dark Matter candidate is a very massive particle, with $m_\chi
\sim 10^{12}$ GeV; decays of this particle could give rise to the
observed very energetic cosmic rays \cite{uhecr} if their lifetime is
$\gsim 10^8$ times the age of the Universe. We noted above that such
massive particles cannot be produced thermally in any realistic model
of inflation. On the other hand, eq.(\ref{phidec}) shows that inflaton
decays might very easily produce too many of such particles. Taking
$m_\phi = 10 m_\chi = 10^{13}$ GeV, we see that we need a branching
ratio as small as $5 \cdot 10^{-8} \ {\rm GeV}/ T_{\rm R} $,
which implies quite a severe upper bound on $T_{\rm R}$ even if $\chi$
pairs can only be produced in six body decays of the inflaton. Even
taking $T_{\rm R} = 1$ MeV, the lowest value compatible with
successful nucleosynthesis, this requires $B(\phi \rightarrow \chi) <
10^{-4}$. Finally, if $\chi$ is produced only through $M_{\rm P}$
suppressed interactions, eq.(\ref{gravbr}) implies $a^2 < 3.5 \cdot
10^{-6} \ {\rm GeV} \cdot M_{\rm P} T_{\rm R} / m^3_\chi$, which again
gives a very tight constraint if $m_\chi \sim 10^{12}$ GeV.

In some cases other considerations give an even stronger constraint on
$\Omega_\chi$. For example, the abundance of charged stable particles
is severely constrained from searches for exotic isotopes in sea water
\cite{ky}, e.g. $\Omega_\chi h^2 \leq 10^{-20}$ for 100 GeV$\lsim
m_\chi \lsim$10 TeV; for heavier particles this bound becomes
weaker. This bound imposes very severe constraints on supersymmetric
models with stable charged LSP. Fixing again $m_\chi = 200$ GeV from
considerations of naturalness, $m_\phi / T_{\rm R} > 4 \cdot 10^{30}
B(\phi \rightarrow \chi)$ is required. This is clearly incompatible
with the limits $T_{\rm R} \gsim 1$ MeV, $m_\phi \lsim 10^{13}$ GeV,
even if $\phi \rightarrow \chi$ decays require six body final states,
see eq.(\ref{sixbody}). We saw above that arranging $\chi$ to have
been in equilibrium at $T_{\rm R}$ does not help. Finally, the relic
density of charged LSPs that were in thermal equilibrium at $T <
T_{\rm R}$ is too large by more than ten orders of magnitude.
Eq.(\ref{phidec}) shows that the situation for larger $m_\chi$ would
be even worse. We thus conclude that in models where at least a
significant fraction of the present entropy of the Universe originates
from inflaton decay, a stable charged LSP can only lead to an
acceptable cosmology if it is too massive to be produced in inflaton
decays.

Our calculation is also applicable to entropy--producing particle
decays that might occur at very late times. If $\chi$ is
lighter than this additional $\phi'$ particle \cite{kamy}, all our
expressions go through with the obvious replacement $\phi \rightarrow
\phi'$ everywhere. More generally our result holds if $\phi$ decays
result in a radiation dominated era with $T_{\rm R} > m_{\phi'}$. If
$\phi'$ is sufficiently long--lived, the Universe will eventually
enter a second matter--dominated epoch. $\phi'$ decays then give rise
to a second epoch of reheating, leading to a radiation--dominated
Universe with final reheating temperature $T_{\rm R_f}$, and
increasing the entropy by a factor $m_{\phi'} / T_{\rm R_f}$. This
could be incorporated into eq.(\ref{phidec}) by replacing $T_{\rm R}
\rightarrow T_{\rm R} T_{\rm R_f} / m_{\phi'} > T_{\rm R_f}$. Our
result regarding a stable charged LSP would remain valid in such a
scenario even if $m_\chi > m_{\phi'}$, since the lower bound of $\sim
1$ MeV which we used now applies to $T_{\rm R_f}$. The only way out
would be to allow $\phi'$ to be essentially the only decay product of
$\phi$, where $\phi'$ itself does not have renormalizable interactions
with standard particles and their superpartners (so that higher order
$\phi$ decays are
negligible) and $2m_\chi > m_{\phi'}$. However, there is presently no
motivation for considering such baroque models.

\section*{Acknowledgements}
This work was supported by
``Sonderforschungsbereich 375 f\"ur Astro-Teilchenphysik'' der
Deutschen Forschungsgemeinschaft.


\end{document}